\documentclass[prb,aps,showpacs,twocolumn,superscriptaddress,nobibnotes,epsf]{revtex4}

\usepackage{graphicx}
\usepackage{dcolumn}
\usepackage{bm}
\usepackage{SIunits}

\begin{document}

\title{Phase Diagram and Calorimetric Properties of NaFe$_{1-x}$Co$_x$As}

\author{A. F. Wang, X. G. Luo, Y. J. Yan, J. J. Ying, Z. J. Xiang, G. J. Ye, P. Cheng, Z. Y. Li, W. J. Hu and X. H. Chen}
\altaffiliation{Corresponding author} \email{chenxh@ustc.edu.cn}
\affiliation{Hefei National Laboratory for Physical Science at
Microscale and Department of Physics, University of Science and
Technology of China, Hefei, Anhui 230026, People's Republic of
China}

\begin{abstract}
We measured the resistivity and magnetic susceptibility to map out
the phase diagram of single crystalline NaFe$_{1-x}$Co$_x$As.
Replacement of Fe by Co suppresses both the structural and magnetic
transition, while enhances the superconducting transition
temperature ($T_{\rm c}$) and superconducting component fraction.
Magnetic susceptibility exhibits temperature-linear dependence in
the high temperatures up to 500 K for all the superconducting
samples, but such behavior suddenly breaks down for the
non-superconducting overdoped crystal, suggesting that the
superconductivity is closely related to the T-linear dependence of
susceptibility. Analysis on the superconducting-state specific heat
for the optimally doped crystal provides strong evidence for a
two-band s-wave order parameter with gap amplitudes of
$\Delta_1(0)/k_{\rm B}T_{\rm c}$= 1.78 and $\Delta_2(0)/k_{\rm
B}T_{\rm c}$=3.11, being consistent with the nodeless gap symmetry
revealed by angle-resolved photoemission spectroscopy experiment.

\end{abstract}

\maketitle

\section{INTRODUCTION}
Superconductivity in the iron-based superconductors, just like that
in cuprates, emerges in the proximity to magnetically ordered state,
so that the magnetic interactions are considered to play a key role
in the mechanism of such high-$T_{\rm c}$ superconductivity.
Accordingly, one of central issues for the iron-based
superconductors is whether SDW follows a local moment or a Fermi
surface nesting picture because this is correlated to the nature of
the pairing force responsible for the
superconductivity.\cite{Hosono, Chen, BaK, Dai PC} The interplay
between magnetism and superconductivity has been extensively
investigated in doped $Ae$Fe$_2$As$_2$ ($Ae$ = Ca, Sr, Ba Eu, and K,
so called "122") family due to easiness of obtaining sizeable and
high-quality single crystals.\cite{BaFe2As2} Phase diagram, through
measurements of electrical transport, magnetism (susceptibility,
$\mu$SR, Neutron scattering, NMR), crystal structure and so on, has
been well studied in such 122 single crystals.\cite{WangXF, neutron,
HuRW, ZhangM, uSR, Imai,HChen} While for the doped NaFeAs system (so
called "111"), studies for phase diagram were performed only on
polycrystalline samples but not on single crystalline ones due to
the hardness to growing high-quality single crystals and the
difficulty to control the doping concentration. According to the
same reason, little work has been done to elucidate the symmetry of
the superconducting gap of doped NaFeAs system
,\cite{Makariy,LiuZH,LiSY} although quite a few studies have been
carried out to explore the superconducting gap structure in the
isostructural LiFeAs by either specific heat or heat transport or
ARPES.\cite{LiFeAshc,Jang,Stockert,LiFeAsht,ARPESLI} Even for these
few studies on the symmetry of the supercondcuting gap of doped
NaFeAs, the results are inconsistent with each other. ARPES and heat
transport experiments revealed nodeless gaps \cite{LiuZH,LiSY} while
the measurements of penetration depth suggested gaps with
nodes.\cite{Makariy}

NaFeAs is established in Fe$_2$As structure with the interstitial Fe
replaced by Na atoms. Therefore, NaFeAs consists of a common
building block, the FeAs layer, and the double layer of Na$^+$
sandwiched between the FeAs layers. NaFeAs is expected to be a
simplified version of the structure of $Re$FeAsO ($Re$ = rare earth,
so called "1111") and "122" iron pnictides. There is no static
magnetic ordering and structural transition in isostructural
compound LiFeAs\cite{LiFeAs,LiFeAsneutron}, which shows
superconductivity at 18 K. However, NaFeAs itself has a
spin-density-wave (SDW) magnetism with a small magnetic moment (0.09
$\mu_{\rm B}$/Fe)\cite{DaiPCNaFeAs}, in contrast to the larger
values of magnetic moment $\sim$ 0.4 $\mu$$_{\rm B}$/Fe in the
La-"1111" \cite{Dai PC} and $\sim$ 0.9 $\mu$$_B$/Fe in the Ba-"122"
parent compounds\cite{neutronBaFe2As2}. Though NaFeAs shows
superconductivity without purposely doping, it possesses only 10\%
superconducting volume fraction and long range AFM order exists in
most of its volume inferred from the $\mu$SR study on the
polycrystalline sample \cite{phasediagram}. By doping with Co on Fe
site, bulk superconductivity in NaFeAs can be achieved. Systematical
study on the doped NaFeAs single crystals by varying Co
concentration has not been done yet. In this paper, we report the
study on the phase diagram of single crystalline
NaFe$_{1-x}$Co$_x$As by measuring resistivity, magnetic
susceptibility and specific heat. Additionally, susceptibility shows
temperature-linear dependence up to 500 K for all the
superconducting samples, and the deviation from high-temperature
linear behavior occurs in the low temperatures, and the deviation
temperature increases slightly with increasing the Co concentration.
However, such behavior is suddenly changed for the
non-superconducting overdoped crystal. Analysis on the specific heat
of the optimally doped sample revealed a two-band s-wave order
symmetry with two gaps with their sizes at $T$ = 0 to be
$\Delta$$_1$(0)/$k_{\rm B}T_{\rm c}$= 1.78 and
$\Delta$$_2$(0)/$k_{\rm B}T_{\rm c}$ = 3.11, suggesting the
strong-coupling superconductivity.

\section{EXPERIMENTAL DETAILS}
High-quality single crystals of NaFe$_{1-x}$Co$_x$As have been grown
by a NaAs flux method. NaAs was obtained by reacting the mixture of
the elemental Na and As in evacuated quartz tubes at 200 $\celsius$
for 10 h. Then Fe, Co and NaAs powder were carefully weighed
according to the ratio of NaAs: Fe: Co=4: 1-$x$: $x$ with ($x$ = 0 -
0.3), and thoroughly ground. The mixtures were put into an alumina
crucible and then sealed in an iron crucible under 1.5 atm of highly
pure argon gas. The sealed crucibles were heated to 950 $\celsius$
at a rate of 60 $\celsius$/h in the tube furnace filled with the
inert atmosphere and kept at 950$\celsius$ for 10 h, and then cooled
slowly to 600 $\celsius$ at 3$\celsius$/h to grow single crystals.
The shiny crystals with typical size of 5$\times$5$\times$0.2 mm$^3$
can be easily cleaved from the melt. X-ray diffraction was perform
on Smartlab-9 diffracmeter ({\sl Rikagu}) from 10$^{\rm o}$ to
60$^{\rm o}$ with a scanning rate of 2$^{\rm o}$ per minute. The
actual chemical composition of the single crystals is determined by
energy dispersive X-ray spectroscopy (EDS). With changing the
nominal $x$ from 0 to 0.3, we obtained 12 batches of crystals, whose
actual compositions were determined by EDS to be 0, 0.006, 0.010,
0.014, 0.017, 0.021, 0.028, 0.042, 0.047, 0.061, 0.070, 0.075 and
0.109, with standard instrument error as about 10\%. Measurements of
resistivity and specific heat were carried out by using the PPMS-9T
({\sl Quantum Design}). Magnetic susceptibility was measured by
using SQUID-MPMS-7T ({\sl Quantum Design}), and a high-temperature
oven was used in the SQUID-MPMS for magnetic susceptibility
measurement above 400 K. In all the magnetic measurements, magnetic
field was applied within ab-plane.

\begin{figure}[ht]
\centering
\includegraphics[width=0.48\textwidth]{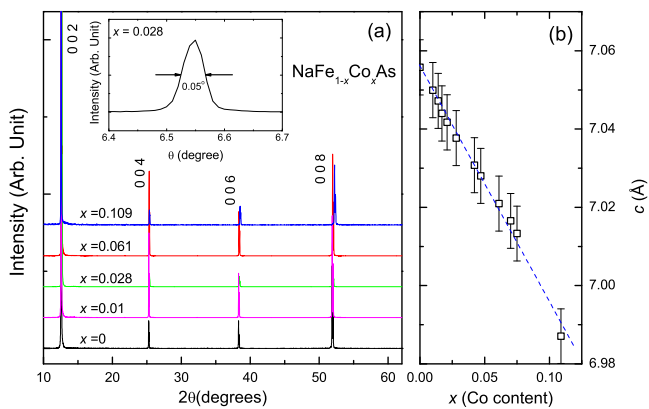}
\caption{(a): Selected XRD patterns for NaFe$_{1-x}$Co$_x$As single
crystals. (b): The lattice parameter of $c$-axis plotted as a
function of Co concentration $x$. Inset in (a) is the rocking curve
for the crystal with $x$ = 0.028.} \label{fig1}
\end{figure}

\begin{figure}[ht]
\centering
\includegraphics[width=0.45\textwidth]{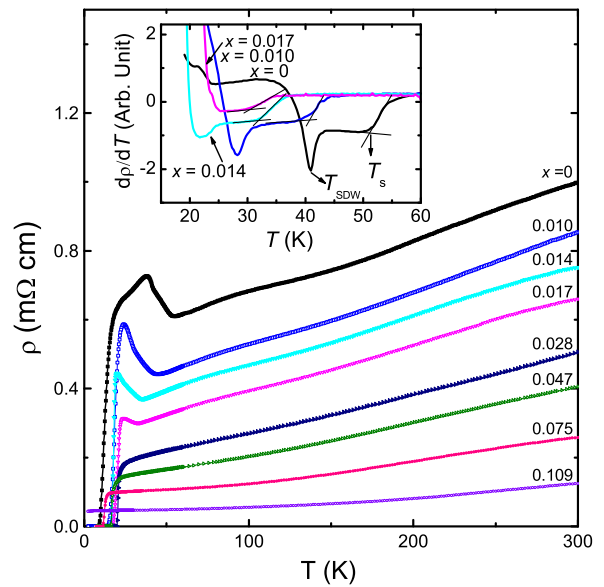}
\caption{Resistivity as a function of temperature for
NaFe$_{1-x}$Co$_x$As single crystals. Inset is the derivative
resistivity for the underdoped NaFe$_{1-x}$Co$_x$As single crystals,
where the scenarios to determine the $T_{\rm s}$ and $T_{\rm SDW}$
are given.} \label{fig2}
\end{figure}

\section{RESULTS AND DISCUSSION}

Figure 1(a) shows the selected single-crystalline XRD patterns for
NaFe$_{1-x}$Co$_x$As single crystals. Only (00$l$) reflections can
be recognized, and the rocking curve plotted in the inset of Fig.
1(a) shows a full-width-at-half-maximum (FWHM) about 0.05$^{\rm o}$,
indicating c-axis orientation and high quality for these single
crystals. Lattice parameter of {\sl c}-axis ($c$) estimated from
Fig. 1(a) was plotted as a function of Co concentration, as shown
Fig. 1(b). The lattice parameter, $c$, of the undoped compound is
7.056 \AA, being consistent with previous reported
results.\cite{phasediagram,Parker} The lattice parameter decreases
linearly with increasing Co doping, consistent with the results
reported on polycrystalline NaFe$_{1-x}$Co$_x$As
samples.\cite{phasediagram} While the amplitude of the change of $c$
($\sim$ 1\% from 0 to 0.109) is much smaller than that reported in
polycrystalline samples ($\sim$ 4\%).\cite{phasediagram}

Temperature dependence of resistivity taken from 2 to 300 K were
displayed in Fig. 2 for the selected single-crystalline
NaFe$_{1-x}$Co$_x$As samples. The resistivity decreases with
increasing Co doping level. An upturn in resistivity is observed in
the low temperatures for the underdoped crystals, which arises from
the structural/SDW transition. The inset shows the derivative of
resistivity to figure out the temperatures corresponding to the
structural and SDW transition ($T_{\rm s}$ and $T_{\rm SDW}$). The
two distinct features in d$\rho(T)$/d$T$ are used to determine the
$T_{\rm s}$ and $T_{\rm SDW}$. The derivative of resistivity is
shown in the inset of Fig. 2. The structural and SDW transitions are
suppressed rapidly with increasing Co concentration. In samples with
Co concentration higher than 2.1\%, no trace of structural/SDW
transition can be recognized in resistivity. The undoped compound is
already superconducting although the superconducting transition is
quite broad, and the resistivity of NaFeAs reaches zero at around 10
K, consistent with previous reports. The superconducting transitions
for most of the samples are quite round, so we define the
temperature at which resistivity reaches zero as $T_{\rm c}$. The
optimal doping level is around $x$ = 0.028 and the corresponding
$T_{\rm c}$ is 20 K. Further Co doping leads to the decrease of
$T_{\rm c}$, and no superconducting transition can be observed down
to 2 K for the crystal with $x$ = 0.109.

\begin{figure}[ht]
\centering
\includegraphics[width=0.45\textwidth]{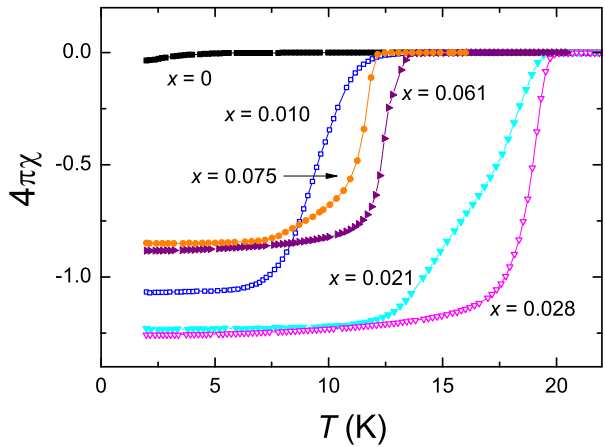}
\caption{Magnetic susceptibility taken at 10 Oe in ZFC mode for the
superconducting samples.} \label{fig3}
\end{figure}

\begin{figure}[ht]
\centering
\includegraphics[width=0.45\textwidth]{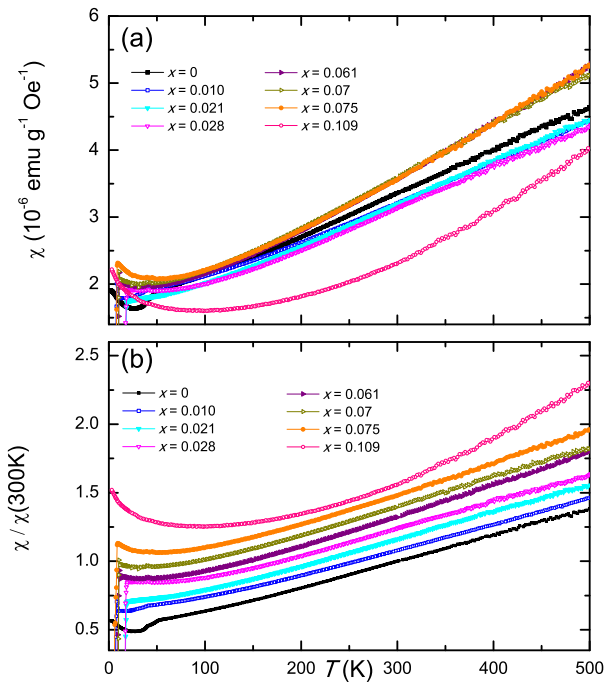}
\caption{(a): Normal-state susceptibility measured at 5 T in the
temperature range up to 500 K for the NaFe$_{1-x}$Co$_x$As single
crystals; (b): Normalized normal-state susceptibility for the
NaFe$_{1-x}$Co$_x$As single crystals. The susceptibility was
normalized to the value at 300 K and shift from sample to sample to
distinguish from each other.} \label{fig4}
\end{figure}

Figure 3 shows the magnetic susceptibility taken at 10 Oe in zero
field cooling (ZFC) procedure for the superconducting
NaFe$_{1-x}$Co$_x$As single crystals. For $x$ = 0, tiny diamagnetic
signal was observed below 9 K, indicating the superconductive
shielding effect is weak in this composition of crystal. With
enhancing Co doping level, $T_{\rm c}$ inferred from diamagnetic
signal increases. The $T_c$ determined by susceptibility is nearly
the same as $T_{\rm c}$ obtained from resistivity measurements for
the crystals with the same composition. The superconductive
shielding fraction rises steeply with increasing Co doping level.
Full shielding at 2 K can be observed for the crystal by
substituting for Fe with 1\% Co. The highest $T_{\rm c}$ of 20 K is
obtained in the crystal with $x$ = 0.028, and the superconducting
transition is very steep. To our knowledge. this is the best for all
the reported crystals in this system. Both shielding fraction and
$T_{\rm c}$ decrease with further doping of Co.

Susceptibility up to 500 K taken under the applied magnetic field of
5 T is shown in Fig. 4(a). Because of the very close magnitude of
the normal-state susceptibilities for the superconducting samples,
the susceptibilities are normalized to the values at 300 K and a
shift was then made for all the Co doped samples to distinguish from
each other, as shown in Fig.4(b). Rapid drops due to superconducting
transition can still be observed at low temperature for the doped
superconducting samples. For the undoped and underdoped samples,
slight kinks can be observed in magnetic susceptibility just above
$T_{\rm c}$ between 20 K and 55 K, which arises from the structural
and SDW transitions. Temperature linear dependence can be observed
in high temperatures for the magnetic susceptibility of all the
superconducting samples. Magnetic susceptibility deviates from the
high-temperature linear behavior in the low temperatures, and the
deviation temperature lies at about 200 K for all the
superconducting crystals. While with further increasing the Co
doping level, only very weak linear behavior can be found above 400
K in heavily overdoped and non-superconducting crystal with $x$ =
0.109. The slope for the linear dependence of high-temperature
susceptibility is the nearly the same for all the superconducting
crystals. Such temperature linear dependence of magnetic
susceptibility is an universal feature for all the iron-based
superconductors, which has been theoretically ascribed to the spin
fluctuations arising from the local SDW correlation.\cite{ZhangGM}
Another theoretical interpretation for the temperature linear
dependence of the susceptibility is taking into account the
peculiarities of orbitally resolved densities of states due to local
correlations, which does not need to invoke the antiferromagnetic
fluctuations.\cite{Klingeler} Whether the temperature linear
dependent susceptibility in high temperatures is correlated with the
pairing force in the high-$T_{\rm c}$ superconductivity of the
iron-based superconducting compounds is still open question. In the
present system, breakdown of the linear dependent susceptibility in
the overdoped region coinciding with disappearance of
superconductivity seems to suggest the association between
temperature linear dependence of the susceptibility and
superconductivity.

\begin{figure}[!ht]
\centering
\includegraphics[width=0.48\textwidth]{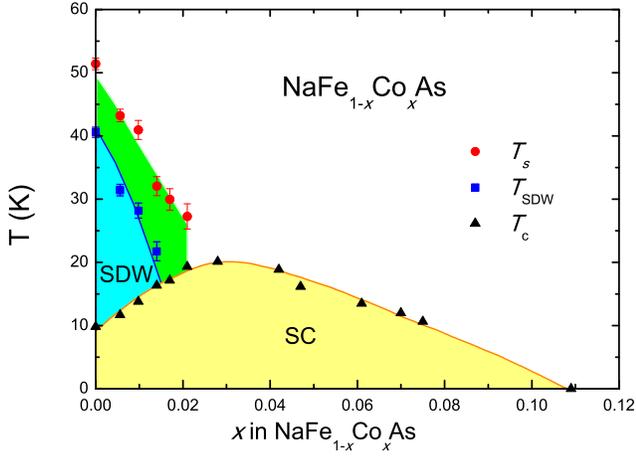}
\caption{Phase diagram of NaFe$_{1-x}$Co$_x$As. $T_{\rm s}$, $T_{\rm
SDW}$ and $T_{\rm c}$ are determined from the resistivity in Fig. 2
and its inset. susceptibility in Fig. 3 and Fig. 4 give almost the
same value for $T_{\rm s}$, $T_{\rm SDW}$ and $T_{\rm c}$.}
\label{fig5}
\end{figure}

Based on $T_{\rm S}$, $T_{\rm SDW}$ and $T_{\rm c}$ inferred from
the resistivity and magnetic susceptibility, the phase diagram of
NaFe$_{1-x}$Co$_x$As system is established from the measurements on
single crystals, as shown in Fig. 5. The gradual destruction of
magnetism and enhancement of superconductivity (rise of $T_{\rm c}$
and superconductive shielding fraction) were observed with
increasing Co doping level. A dome-shaped $T_{\rm c}$ vs. $x$
relationship can be observed. Optimal $T_{\rm c}$ was obtained
around $x$ = 0.028 and further Co doping suppressed
superconductivity. $T_{\rm c}$ goes to zero around $x$ = 0.109. This
phase diagram is quite similar to those of 122 and 1111 systems
except for that the starting compound of the present system is
superconducting. In addition, the optimal $T_{\rm c}$ can be
achieved by about 2.8\% Co doping in NaFe$_{1-x}$Co$_x$As, much less
than $\sim$7\% Co in Ba(Fe$_{1-x}$Co$_{x}$)$_2$As$_2$
system.\cite{NiNi,Fisher} From this point of view, the phase diagram
of NaFe$_{1-x}$Co$_x$As is more similar to that of
Ba(Fe$_{1-x}$Ni$_{x}$)$_2$As$_2$ system,\cite{NiNi1} in which the
optimal doping is reached at 4.6\%, far from that of $\sim$7\% in
Ba(Fe$_{1-x}$Co$_{x}$)$_2$As$_2$.

\begin{figure}[!htp]
\centering
\includegraphics[width=0.48\textwidth]{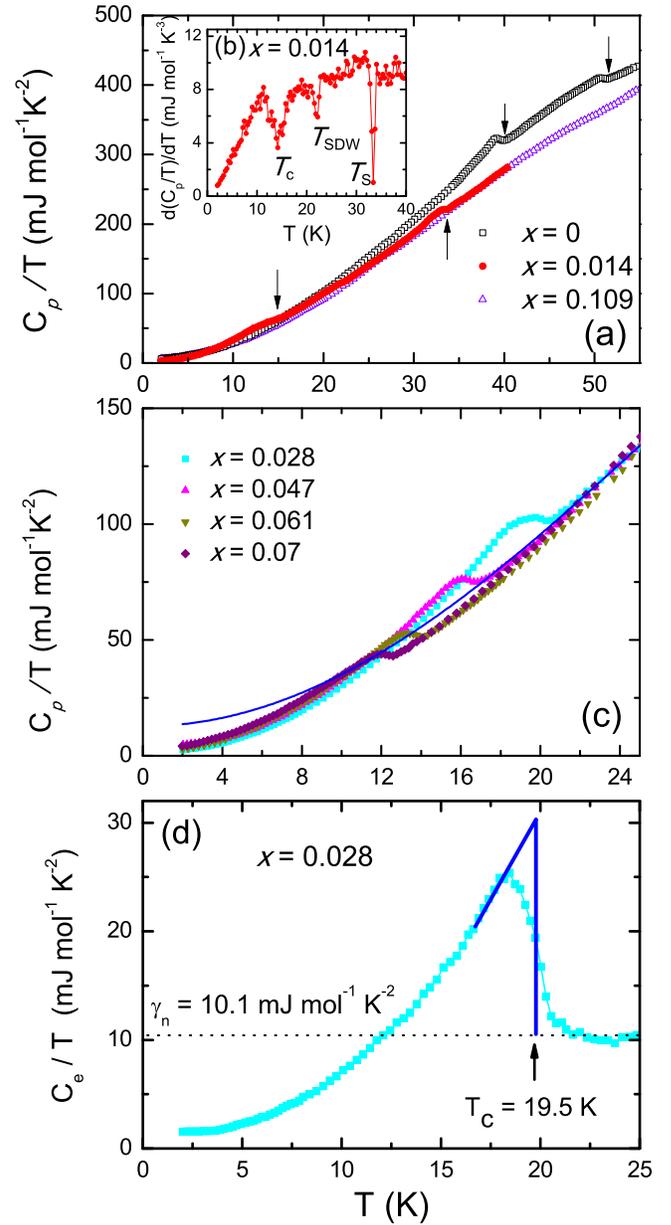}
\caption{(a): Temperature dependence of low-temperature specific
heat ($C_p/T$) for underdoped and heavily overdoped
NaFe$_{1-x}$Co$_x$As crystals. The arrows points to the anomalies in
the specific heat. (b): The inset  shows the derivative of specific
heat for the crystal with $x$ = 0.014, where structural, SDW and
superconducting transitions can be clearly recognized from the dips.
(c): Specific heat as a function of temperature for the
optimally-doped and overdoped superconducting NaFe$_{1-x}$Co$_x$As
crystals. The blue line is the fitting of the specific heat between
20 - 40 K by using $C_p$ = $C_{en}$+$C_{lattice}$. (d): Electronic
specific heat $C_e$/$T$ (by subtracting the lattice contribution
from $C_p/T$) as function of temperature for the optimally doped
crystal, where the dashed line represents the normal-state
electronic contribution, $\gamma_n$ = 10.1 mJ mol$^{-1}$ K$^{-2}$.}
\label{fig6}
\end{figure}

Figure 6 displays the temperature dependence of the low-temperature
specific heat (plotted as $C_p/T$) for the underdoped and heavily
overdoped crystals. The specific heat for the undoped samples shown
in Fig. 6(a) exhibit two anomalies corresponding to the structural
and SDW transitions, respectively. The $T_{\rm S}$ and $T_{\rm SDW}$
are well consistent with those determined from resistivity in the
inset of Fig. 2. No anomaly corresponding to the superconducting
transition can be observed in the specific heat for the undoped
sample, which could be due to the low fraction of superconducting
component. This is consistent with the tiny superconducting
shielding fraction from susceptibility measurement shown in Fig. 3.
For the crystal with $x$ = 0.014, a clear anomaly corresponding to
the supercondcuting transition around 14 K, indicating bulk
superconductivity in this sample and consistent with the good
superconductive shielding for crystals with $x >$ 0.01 (as shown
Fig. 3). In addition, an anomaly is observed at about 33 K, which
coincides with the structural transition determined by the
derivative of resistivity shown in the inset of Fig.2. However, no
clear anomaly corresponding to SDW transition is observed. In order
to distinguish from the structural, SDW and superconducting
transitions, the derivative of specific heat is shown in Fig.6(b),
and Fig.6(b) clearly shows three dips at $T_{\rm S}$, $T_{\rm SDW}$
and $T_{\rm c}$, indicating the coexistence of the
antiferromagnetism and superconductivity for the underdoped crystal
with x=0.014. The specific heat of the heavily overdoped and
non-superconducting crystal with $x$ = 0.109 does not show any
anomaly. As temperature increases, the specific heat above $T_{\rm
s}$ for the sample with $x$ = 0.014 tends to be the same as that of
the sample with $x$ = 0.109.

\begin{figure}[ht]
\centering
\includegraphics[width=0.48\textwidth]{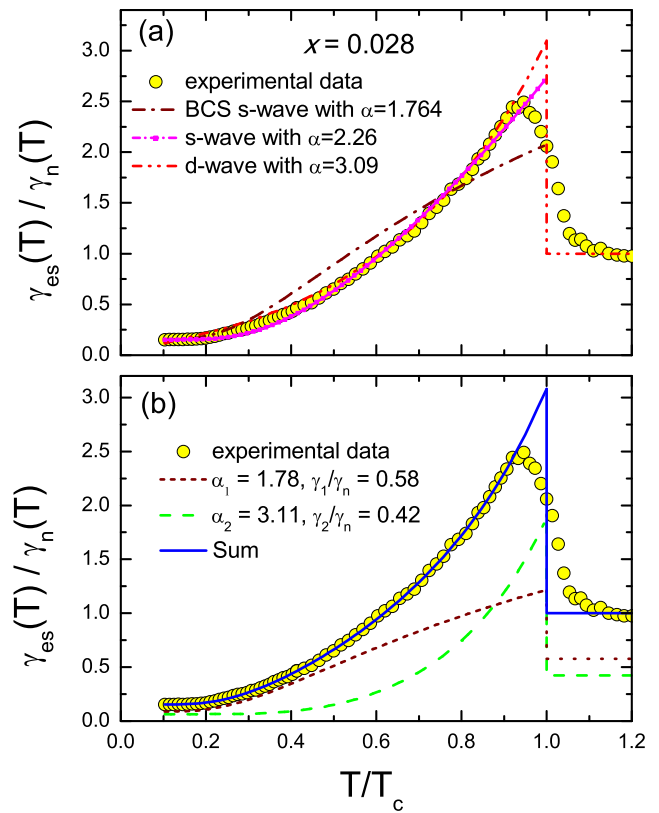}
\caption{(a): The normalized electron specific heat of the optimally
doped crystal with $x$=0.028, compared with the specific heat of
single-band s-wave (wine line) and d-wave (magenta line) order
parameters as well as that in the weak-coupling limit (red line).
(b): The normalized electron specific heat of the superconducting
sample with $x$=0.028. The blue curve represents a two-gap fit. The
wine and green curves are the partial specific-heat contributions of
the two bands.} \label{fig7}
\end{figure}

Pronounced jump due to superconducting transition can be observed in
the temperature dependence of specific heat for the optimally-doped
and overdoped superconducting samples, as shown in Fig. 6(c). The
normal-state specific heat, $C_p$, can be described by $C_p$ =
$C_{en}$ + $C_{\rm lattice}(T)$ with the electronic contribution of
$C_{en}$ = $\gamma_{n}T$  and  the lattice contribution of $C_{\rm
lattice}(T)$ = $\beta T^3$ + $\delta T^5$. The solid line in Fig.
6(c) is the best fit to the $C_p/T$ above $T_{\rm c}$ (20 K -40 K)
for the optimally doped sample, yielding $\gamma_{n}$ = 10.1 mJ/mol
K$^{-2}$, $\beta$ = 0.23 mJ/mol K$^{-4}$ and $\delta$ = -0.0589
$\mu$J/mol K$^{-6}$. After subtracting this $C_{\rm lattice}$
component, the obtained results are shown in Fig. 6(d), for which
the entropy conservation is confirmed to be satisfied. Figure 6(d)
shows that $C_e/T$ does not extrapolate to zero at $T$ = 0 limit but
to a finite residual normal-state-like contribution $\gamma_r$ = 1.5
mJ mol$^{-1}$ K$^{-2}$. Finite values of $\gamma_r$ are a common
feature in the specific measurements of iron-based
superconductors.\cite{Hardy}

Normalized electron specific heat is displayed in Fig. 7, where
$\gamma_{es}$ = $C_e/T$. Figure 7(a) indicates that $C_e$ cannot be
described by a single band BCS superconductor, calculated either in
the weak-coupling limit with  $\alpha$ = $\Delta(0)$/$k_{\rm
B}$$T_{\rm c}$ = 1.764 or by letting $\alpha$ adjustable.
Especially, the fitting in weak-coupling limit shows very poor
agreement with the experimental data. Figure 7(a) also shows the
fitting results by using the single-band d-wave model. It is clear
that such a $k$-dependent gap, even in strong coupling scenario
(obtained $\alpha$ = 3.09), cannot describe the observed data. As a
result, the phenomenological two-band $\alpha$ model with two energy
gaps was used, and the fitting results are shown in Fig. 7(b). The
two-band $\alpha$ model results in an excellent fitting of the
specific heat from $T_{\rm c}$/10 to $T_{\rm c}$, and as a
consequence gives reliable gaps. This fitting is calculated as the
sum of the contributions from two bands by assuming independent BCS
temperature dependences of the two superconducting gaps, as shown by
the dashed and short-dashed lines in Fig. 7(b). In the fitting, the
magnitudes of two gaps at $T$ = 0 limit are introduced as adjustable
parameters, $\alpha_1$ = $\Delta_1(0)$/$k_{\rm B}$$T_{\rm c}$ and
$\alpha_2$ = $\Delta_2(0)$/$k_{\rm B}$$T_{\rm c}$. At the same time,
$\gamma_i/\gamma_n$ ($i$ = 1, 2), which measures the fraction of the
total normal electron density of states, are introduced as
adjustable parameters. The obtained $\alpha_1$ for small gap is
1.78, and $\alpha_2$ for large gap is 3.11. The relative weight for
the small and large gaps here is $\gamma_1$/$\gamma_2$ = 0.72. Both
of the obtained two gaps are larger than that of the BCS
weak-coupling limit. This is not consistent with the theoretical
constraints that one gap must be larger than the BCS gap and one
smaller in a weakly coupled two-band superconductor,\cite{Kresin}
which has been observed in MgB$_2$ \cite{MgB2} and another
iron-based superconductor Ba(Fe$_{0.925}$Co$_{0.075}$)$_2$As$_2$
~\cite{Hardy} and as well as another 111 superconductor
LiFeAs.\cite{Jang} Actually, such phenomenon that the two gaps are
larger than the BCS weak-coupling limit has been also observed in
Fe(Te$_{0.57}$Se$_{0.43}$) ~\cite{FeTeSe} and
Ba$_{0.6}$K$_{0.4}$Fe$_2$As$_2$ ~\cite {BaK0p4HC} single crystals.
The large magnitudes of gaps derived from our fitting suggest the
strong-coupling superconductivity, as that reported in
Fe(Te$_{0.57}$Se$_{0.43}$) previously.\cite{FeTeSe} The relative
ratio of the two gaps in NaFe$_{0.972}$Co$_{0.028}$ is
$\Delta_1(0)$/$\Delta_2(0)$ $\sim$ 0.57, which is a little larger
than that seen in other iron-pnictide superconductors where
$\Delta_1(0)$/$\Delta_2(0)$ = 0.3 $\sim$
0.5.\cite{Gofryk,Hardy,Hardy1,Popovich} It should be pointed out
that although in the present work the fitting of specific heat
suggests isotropic gaps, some theoretical and other experiments also
suggest complicated pair symmetry for most of iron-based
superconductors, including the nodeless anisotropic or nodal gaps.
Actually, penetration depth experiment indicates a nodal gap in the
optimally doped crystal,\cite{Makariy} while ARPES and heat
transport measurements revealed nodeless gaps in the optimally doped
and overdoped NaFe$_{1-x}$Co$_x$As crystal,\cite{LiuZH,LiSY} Our
results supports the nodeless gap symmetry observed by ARPES
experiment.\cite{LiuZH} Especially, our analysis is consistent with
the existence of different magnitudes of the gaps suggested by heat
transport.\cite{LiSY}

\section{CONCLUSION}
In summary, we mapped out the phase diagram of NaFe$_{1-x}$Co$_x$As
system through measuring transport and magnetic properties on
high-quality single crystals. Substitution of Co for Fe destroys
both the structural and magnetic transition, while enhances the
superconducting transition temperature ($T_{\rm c}$) and
superconducting shielding fraction. Susceptibility exhibits
temperature-linear dependence in the high temperatures up to 500 K
for all the superconducting samples, while such behavior cannot be
observed in the non-superconducting overdoped crystal. The breakdown
of the linear dependent susceptibility in the overdoped region
coinciding with disappearance of superconductivity seems to suggest
the association between temperature linear dependence of the
susceptibility and superconductivity. Analysis on the specific heat
in superconducting state reveals a two-band s-wave order parameter
with gap amplitudes of $\Delta_1(0)/k_{\rm B}T_{\rm c}$= 1.78 and
$\Delta_2(0)/k_{\rm B}T_{\rm c}$=3.11, suggesting the
strong-coupling superconductivity.

\begin{acknowledgments}
This work is supported by National Natural Science Foundation of
China (Grant No. 11190021, 51021091), the National Basic Research
Program of China (973 Program, Grant No. 2012CB922002 and No.
2011CB00101), and Chinese Academy of Sciences.

\end{acknowledgments}

\end{document}